\documentclass[journal=jpcbfk,manuscript=article]{achemso}
\usepackage[version=3]{mhchem}  
\usepackage{times,mathptmx}
\usepackage[english]{babel}
\usepackage[utf8]{inputenc}
\usepackage{amsmath}
\usepackage{amssymb}
\usepackage{amsthm}
\usepackage{graphicx}
\usepackage{bm}
\usepackage{bbm}
\usepackage{empheq}
\usepackage{verbatim}
\usepackage{graphicx}
\graphicspath{{Images/}{../Images/}}

\usepackage{color}
\usepackage{pagecolor}
\pagecolor{white}

\newcommand{\changetextcolor}{\color{black}}
\newcommand{\imagewidth}{0.6\columnwidth}
\newcommand{\f}{\frac}
\newcommand{\mm}[1]{\mathrm{#1}}
\newcommand{\vr}{\bm{r}}
\newcommand{\vecr}[1]{\bm{r}#1}
\newcommand{\dvr}[1]{\mathrm{d}\vecr{#1}\,}
\newcommand{\ds}{\mathrm{d}s\,}

\newcommand{\bc}{bicontinuous}
\newcommand{\bcp}{block copolymer}
\newcommand{\dbc}{diblock copolymer}
\newcommand{\hp}{homopolymer}
\newcommand{\hpl}{homopolymer-like}
\newcommand{\dg}{double gyroid}
\newcommand{\dd}{double diamond}
\newcommand{\pn}{plumber's nightmare}
\newcommand{\mcurv}{mean curvature}
\newcommand{\cmc}{constant \mcurv{}}

\newcommand{\scft}{SCFT}
\newcommand{\adbc}{DBCP}
\newcommand{\ahp}{H}

\newcommand{\DIS}{DIS} 
\newcommand{\LAM}{L} 
\newcommand{\HEX}{C} 
\newcommand{\DD}{DD}
\newcommand{\DG}{G}
\newcommand{\PN}{P} 

\newcommand{\imon}{\gamma}
\newcommand{\icha}{c}
\newcommand{\ig}{\mm{G}}
\newcommand{\ihl}{\mm{HL}}
\newcommand{\ihp}{\mm{H}}

\newcommand{\vmon}{v} 
\newcommand{\kuhnl}{b} 
\newcommand{\xAB}{\chi} 
\newcommand{\segstr}{\xAB N^{\mm{\ig}}} 
\newcommand{\fA}[1]{f_{\mm{A},#1}} 
\newcommand{\fhl}{f_{\mm{A},\ihl}}
\newcommand{\fdg}{f_{\mm{A},\ig}}
\newcommand{\alphahp}{\alpha_\ihp}
\newcommand{\avgpc}[1]{\overline{\phi}_{#1}} 

\newcommand{\vfhl}{\avgpc{\ihl}}
\newcommand{\vfhp}{\avgpc{\ihp}}

\newcommand{\NA}[1]{N_{\mm{A},#1}}  
\newcommand{\Nt}[1]{N_{#1}} 

\newcommand{\fint}{\int\mathcal{D}} 
\newcommand{\pA}[1]{\phi_{\mm{A}}(\vecr{#1})}
\newcommand{\pB}[1]{\phi_{\mm{B}}(\vecr{#1})}
\newcommand{\pC}[1]{\phi_{\mm{#1}}(\vecr)}

\newcommand{\wA}[1]{\omega_{\mm{A}}(\vecr{#1})}
\newcommand{\wB}[1]{\omega_{\mm{B}}(\vecr{#1})}
\newcommand{\wC}[1]{\omega_{\mm{#1}}(\vecr)}
\newcommand{\xif}[1]{\Xi(\vecr{#1})} 
\newcommand{\zpot}[1]{e^{\mu_{#1}}} 

\title[]{\changetextcolor Binary Blends of Diblock Copolymers: An Effective Route to Novel Bicontinuous Phases}

\author{\changetextcolor Chi To Lai}
\email{laic13@mcmaster.ca}
\affiliation{\changetextcolor  Department of Physics \& Astronomy, McMaster University, 1280 Main St. W, Hamilton, Ontario L8S 4M1, Canada}

\author{\changetextcolor  An-Chang Shi}
\email{shi@mcmaster.ca}
\affiliation{\changetextcolor Department of Physics \& Astronomy, McMaster University, 1280 Main St. W, Hamilton, Ontario L8S 4M1, Canada}

\begin{tocentry}
Various bicontinuous phases are stabilized in diblock copolymer blends
\centering
\includegraphics[height=4.0cm]{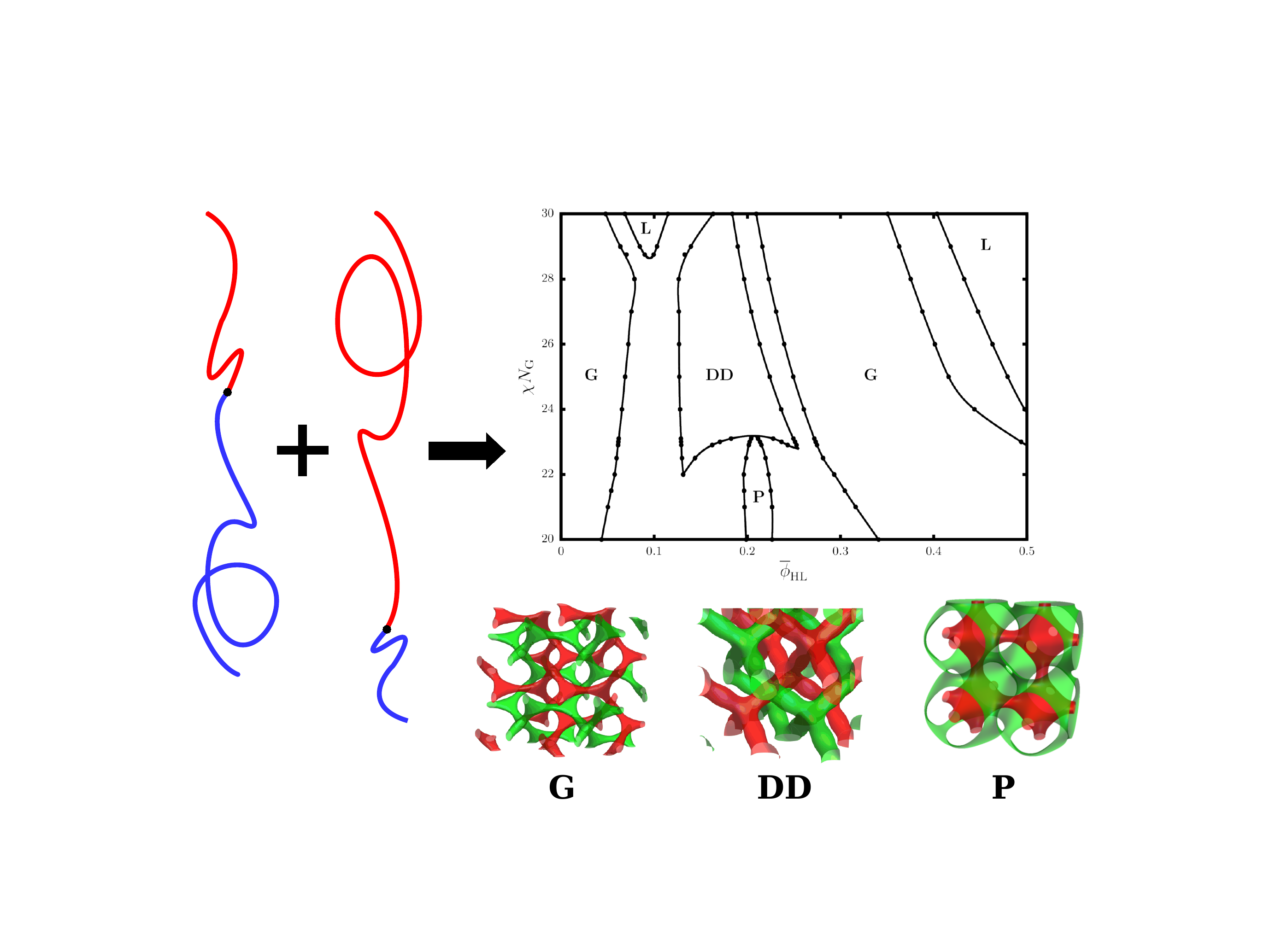} 
\vspace{2cm}
\end{tocentry}

\begin{document}
\begin{abstract}
\changetextcolor 
The formation of various \bc{} phases from binary blends of linear AB diblock copolymers (\adbc{s}) is studied using the polymeric self-consistent field theory. The theoretical study predicts that the double-diamond and the ``plumber's nightmare'' phases, which are metastable for neat diblock copolymers, could be stabilized in block copolymers with designed dispersity, namely, binary blends composed of a gyroid-forming \adbc{} and a homopolymer-like \adbc{}. The spatial distribution of different monomers reveals that these two types of \adbc{s} are segregated such that the homopolymer-like component is localized at the nodes to relieve the packing frustration. Simultaneously, the presence of a local segregation of the two \adbc{s} on the AB interface regulates the interfacial curvature. These two mechanisms could act in tandem for homopolymer-like \dbc{}s with proper compositions, resulting in larger stability regions for the novel \bc{} phases.
\end{abstract}


\changetextcolor 
\section{Introduction}
\changetextcolor	
The most distinguishing property of \bcp{}s is their ability to self-assemble into periodically ordered structures with periods on the order of 10-100 nm~\cite{hamley1998physics}. In the simplest case of linear AB \dbc{}s, the morphology of the self-assembled structure could be systemically regulated by varying a small number of parameters. Two of these parameters are the volume fraction of the A-blocks, $f_A$, and the segregation strength $\chi N$, where $\chi$ is the Flory-Huggins parameter quantifying the immiscibility of the two unlike monomers and $N$ denotes the total polymer length~\cite{hamley1998physics}. Recently, it has also been recognized that the conformational asymmetry of block copolymers could play an important role in stabilizing complex spherical packing phases~\cite{Li:2017bo}. Starting from symmetric \dbc{}s with $f_A = 0.5$, the ordered phases of the system follow a generic transition sequence of lamellar (L) $\rightarrow$ cylindrical (C) $\rightarrow$ \bc{} (G) $\rightarrow$ spherical (S) phases as the \adbc{s} become more asymmetric~\cite{hamley1998physics}. Among the various ordered phases of block copolymers, the \bc{} phases are of particular interest due to their intricate structures and their potential applications in the production of highly-porous materials \cite{kakiage06}, materials with low refractive index~\cite{Hsueh2010}, high-conductivity nanoparticles \cite{Cho1598}, dye-sensitive solar cells \cite{jhenry2009}, and 3D photonic crystals \cite{elthomas2002}. Therefore, it is desirable to search for polymeric systems that could form various bicontinuous phases in a large phase space.

It is worthwhile to first briefly review the discovery of the \bc{} phases self-assembled from \bcp{}s. The first report of a \bc{} structure in AB-type \bcp{}s, initially designated as the double-diamond (\DD{}) phase, was made in 1986 for poly-(styrene-\textit{b}-isoprene) (SI) star copolymers~\cite{alward1986effect,thomas1986ordered}. A year later, Hashimoto {\em et al.}~\cite{hasegawa1987bicontinuous} observed a four-branched tetrapod-network structure in linear SI \dbc{}s, which was regarded as an example of the \dd{} structure. In 1994, Hajduk {\em et al.} reported the formation of the double gyroid (\DG{}) phase in star copolymers~\cite{hajduk1994gyroid} and in linear \dbc{}s~\cite{hajduk1995reevaluation}. In these two studies, the authors pointed out that the bicontinuous structures observed in previous reports~\cite{alward1986effect,thomas1986ordered} would be more compatible with the double gyroid phase. At the same time,  Schulz {\em et al.}~\cite{schulz1994epitaxial} also observed the \dg{} phase in poly(isoprene-\textit{b}-2-vinylpyridnie). Shortly after, a self-consistent field theory calculation carried out by Matsen and Schick~\cite{Matsen1994} predicted that the \DG{} phase has lower free energy than that of the \DD{} phase, thus confirming the designation of Hajduk {\em et al.} and Schulz {\em et al.}. It is now generally accepted the only bicontinous structure accessible in monodisperse melts of linear AB \dbc{}s is the \dg{} phase~\cite{meuler2009ordered}. 

There exist a number of bicontinuous networked structures, consisting of two interweaving networks composed of the minority-block monomers embedded in a matrix of the majority block~\cite{meuler2009ordered,fsbates1996}. The networks could be viewed as struts or connectors joining together at different nodes. The number of struts meeting at a node depends on the particular structure. For the \dg{} ($Ia\overline{3}d$), the \dd{} ($Pn3m$), and the plumber's nightmare (\PN) ($Im\overline{3}m$) phases, the number of struts per node is 3, 4, and 6, respectively. 
In order to minimize the surface area of the AB interface under the constant-volume constraint, the formation of a constant-mean-curvature structure is preferred~\cite{Matsen:2002p437}. For the \bc{} phases, this leads the nodes becoming thicker or bulkier than the connectors, with the disparity in size increasing with the number of struts per nodes~\cite{fsbates1996}. On the other hand, in order to maintain a uniform monomer density, polymers must stretch excessively to fill the space at the center of the nodes~\cite{Matsen:2002p437,hasegawa1996}. The competition of these two opposing trends results in packing frustration in the system. It is argued that the stability of the \dg{} phase in monodisperse \dbc{}s originates from the fact that the thickness difference between the nodes and struts is the smallest in the \dg{} structure, thus resulting in the least packing frustration~\cite{Matsen:2002p437}.

One way to alleviate the packing frustration at the nodes is to employ polymeric systems with designed dispersity distributions. It is well known that dispersity of block copolymers could be used to regulate their phase behaviour~\cite{fredrickson04cont,shi06effects,Matsen:2007p1286,Lynd:2008p1906}. 
The simplest disperse distribution is bidisperse systems, {\it viz.} binary blends, composed of mixtures of two types of polymers, which exhibit rich phase behaviour and offer opportunities to form desirable structures~\cite{Liu:2016ky}.
For the case of bicontinous phases, Matsen showed theoretically that the \dd{} phase could be accessed in \hp{}-\dbc{} (\ahp{}+\adbc{}) blends~\cite{matsen1995stabilizing}. Specifically, his self-consistent field theory (SCFT) calculations demonstrated that the added homopolymers localize at the centers of the nodes, which reduces the need for the \dbc{}s to stretch excessively to fill the space, thus relieving packing frustration. The approach of introducing space-fillers was further pursued by Escobedo and coworkers using a number of theoretical techniques~\cite{Martinez-Veracoechea2007,Martinez-Veracoechea2009,Martinez-Veracoechea2009_PLNM, Martinez-Veracoechea2016}.  In a Monte Carlo simulation study, Mart{\'{i}}nez-Veracoechea and Escobedo~\cite{Martinez-Veracoechea2007} considered selective solvents particles in addition to homopolymers as the additive component. The solvent particles were found to be distributed uniformly throughout the minority domain, and therefore, failed to reduce the packing frustration that is necessary to stabilize the \dd{}, or \pn{} phases. Using \scft{}, Mart{\'{i}}nez-Veracoechea and Escobedo~\cite{Martinez-Veracoechea2009_PLNM} further demonstrated that both the blend composition or the \hp{} length could be tuned to stabilize the \pn{} phase in \ahp{}+\adbc{} blends. A subsequent molecular dynamics study~\cite{Martinez-Veracoechea2016} yielded information on the chain conformations within the \bc{} structures, which provided direct evidence that the addition of \hp{}s alleviates the packing frustration occurring at the nodes. Experimentally, Takagi {\em et al.}~\cite{takagi2015ordered} probed the phase behaviour of binary blends of SI \dbc{}s and polyisoprene \hp{}s using small angle x-ray scattering. They observed the formation of the \dd{} phase, and moreover found that the order-order transition between the \DD{}, and \DG{} phases was thermodynamically reversible. Most recently, Takagi {\em et al.}~\cite{takagi2019bicontinuous} demonstrated that the \dd{} phase can be accessed in ternary blends of SI \dbc{}s at specific blend compositions.

To the best of our knowledge, existing theoretical studies examining the formation of \bc{} phases in \dbc{} blends have either considered \hp{}s or solvent particles as space-filling additives. One limitation of using \hp{}s is that macrophase separation occurs between the \hp{}-rich disordered phase and the adjacent \bc{} phase at relatively low homopolymer concentrations, resulting in a small range of blend compositions over which the \dd{} or \pn{} phases could be formed. A possible strategy to circumvent macrophase separation is to use a polydisperse \dbc{}s in which the long and short chains could segregate to relieve the packing frustration. In this paper, we examine a bidisperse system composed of two types of AB diblock copolymers, namely a gyroid-forming \dbc{} and a homopolymer-like diblock copolymer, where the A-blocks are taken as the network-forming component. The homopolymer-like AB diblock copolymer has an A-block that is much longer than its B-block. We employ the polymeric \scft{} to construct the phase diagrams of the system. The effects of the blend composition and the volume fraction of the \dbc{}s are examined. We find that both the \dd{} and \pn{} phases could emerge as stable phases when the total length of the \hpl{} component becomes sufficiently long. Moreover, the range of blend compositions over which the \dd{} or \pn{} phases occurs is found to be larger than that of the equivalent \ahp{}+\adbc{} blends. By examining the spatial distribution of the polymeric species within the self-assembled structures, we find evidence that the \hpl{} copolymers not only act as a space-filler that relieves packing frustration, but also act as co-surfactants resulting in favourable changes to the AB-interfacial curvature. The combination of these two effects results in the larger stable regions of the various bicontinuous phases.

\section{Theoretical Model}
\label{sec_background}
We consider a binary mixture of \dbc{}s, (AB)$_1$ and (AB)$_2$, in a volume $V$. The length of the A-block, and the total length of the copolymers are given by $N_\mm{A,c}$, and $N_c$, respectively with the index $c=1$ or $2$ labeling the type of diblock copolymers. The melt is assumed to be incompressible. We also assume for simplicity that the A-, and B-monomer volumes, as well as their Kuhn lengths are equal, {\em i.e.} $\vmon_\mm{A} = \vmon_\mm{B}$, and $\kuhnl_\mm{A} = \kuhnl_\mm{B}$. The polymer chains are modeled as Gaussian chains~\cite{fredrickson2006equilibrium}.

Information about he thermodynamics of the system is contained in the partition function, which can be expressed as a functional integral over all chain conformations. Alternatively, the partition function can be expressed in the grand canonical ensemble as a functional integral over the monomer volume fractions 
$\phi_\imon(\vr{})$ and auxiliary fields $\omega_\imon(\vr{})$ as,
\begin{equation}
\begin{aligned}
Z =& \fint \pA{} \fint \pB{} \fint \wA{} \fint \wB{} \fint \Xi(\vr{})
\\ 
&\times \exp\left\{-\beta G[\pA{},\wA{},\pB{},\wB{},\Xi(\vr{})]\right\},
\end{aligned}
\end{equation}
	where $G=G[\pA{},\wA{},\pB{},\wB{},\Xi(\vr{})]$ is the free energy functional given in units of $k_B T$,~\cite{fredrickson2006equilibrium,shi2017variational}
\begin{equation}
\label{eq_fed_scft}
\begin{aligned}
\f{G}{\rho_0 V}  =& \f{1}{V}\bigg\{\int\dvr{} \xAB \pA{} \pB{} - \sum_{\imon=\mm{A},\mm{B}} \wC{\imon}\pC{\imon} 
\\
& - \Xi(\vecr{})\left[1 - \pA{} - \pB{}\right]\bigg\} - \sum^{2}_{\icha=1} \zpot{\icha}Q_{\icha}.
\end{aligned}
\end{equation}
Here, the index $\imon$ runs over the monomer types ({\em i.e.} $\imon =$  A, B), and $\mu_\icha$ is the chemical potential of $\icha$-type polymers in units of $k_B T$. Due to the incompressibility condition, the two chemical potentials are not independent. We can therefore set one of the chemical potentials to a convenient reference value, and vary the other in order to adjust the blend  composition~\cite{matsen1995phase}. Moreover, we have introduced a Lagrange multiplier $\Xi(\vr{})$ to enforce the incompressibility condition. The single-chain partition function for $\icha$-type polymers is defined by,
\begin{equation}
Q_{\icha} = \f{1}{V}\int\dvr{} q_{\icha}(\vecr{},\Nt{\icha}),
\end{equation} 
where $q_{\icha}(\vecr{},\Nt{\icha})$ is the forward end-integrated propagator obeying the modified diffusion equation, 
\begin{equation}
\label{eq_mod_diff_p1}
\f{\partial q_{\icha}(\vecr{},s)}{\partial s} = \nabla^2 q_{\icha}(\vecr{},s) - \omega_{\icha}(\vecr{}, s) q_{\icha}(\vecr{},s),
\end{equation}
\begin{equation}
\label{eq_mod_diff_p2}
\begin{aligned}
\omega_{\icha}(\vecr{},s) = \bigg\{
\begin{array}{cc}
\wA{} & 0\leq s \leq \NA{\icha} \\
\wB{} &  \NA{\icha} \leq s \leq \Nt{\icha} 
\end{array}
\end{aligned},
\end{equation}
subject to the initial condition $q_{\icha}(\vecr{},0)= 1$. 

In order to evaluate the free energy of the system, we apply the mean-field approximation~\citep{fredrickson2006equilibrium,shi2017variational}, which amounts to minimizing the free energy density functional with respect the monomer concentrationss, the auxiliary fields, and the Lagrange multipliers. 
The mean-field approximation leads to a set of SCFT equations whose solutions corresponding to various ordered phases of block copolymers could be obtained using available numerical methods. It is also noted that the mean-field approximation also ignores correlations between fluctuations that could be important near the critical point of the system~\citep{fredrickson2006equilibrium}.
The minimization procedure leads a set of self-consistent equations. Performing the minimization with respect to the concentrations gives,
\begin{equation}
\label{eq_omega_scft}
\wA{} = \xAB \pB{} + \xif{}, \quad \wB{} = \xAB \pA{} + \xif{}.
\end{equation}
The incompressibility condition is recovered when minimizing with respect to $\xif{}$,
\begin{equation}
\label{eq_incompress_cond}
\pA{} + \pB{} = 1.
\end{equation}
We can lastly determine the equilibrium monomer volume fractions by extremizing with respect to the auxiliary fields, which yields,
\begin{equation}
\label{eq_phi_scft}
\begin{aligned}
\pA{} = \sum_{i=1}^{2}\zpot{\icha}\int^{\NA{\icha}}_0\ds q_{\icha}(\vecr{},s) q_{\icha}^{\dagger}(\vecr{},\Nt{\icha}-s),
\\
\pB{} = \sum_{i=1}^{2}\zpot{\icha}\int^{\Nt{\icha}}_{\NA{\icha}}\ds q_{\icha}(\vecr{},s) q_{\icha}^{\dagger}(\vecr{},\Nt{\icha}-s),
\end{aligned}
\end{equation}
where $q_{\icha}^{\dagger}(\vecr{},s)$  is the backward end-integrated propagator that satisfies a similar diffusion equation to Eq.~\eqref{eq_mod_diff_p1},
\begin{equation}
\label{eq_mod_diff_p3}
\f{\partial q^{\dagger}_{\icha}(\vecr{},s)}{\partial s} = \nabla^2 q^{\dagger}_{\icha}(\vecr{},s) - \omega^{\dagger}_{\icha}(\vecr{}, s) q_{\icha}(\vecr{},s),
\end{equation}
\begin{equation}
\label{eq_mod_diff_p4}
\begin{aligned}
\omega^{\dagger}_{\icha}(\vecr{},s) = \bigg\{
\begin{array}{cc}
\wB{} & 0\leq s \leq \Nt{\icha}-\NA{\icha} 
\\
\wA{} &  \Nt{\icha}-\NA{\icha} < s \leq \Nt{\icha} 
\end{array}
\end{aligned},
\end{equation}
once more subject to the initial condition, $ q^{\dagger}_{\icha}(\vr,0) = 1$.

There are multiple solutions to the self-consistent equations, corresponding to different possible phases of the system. Spatially-varying solutions correspond to the ordered structures, while a constant solution that can always be found represents the homogeneous disordered phase. Barring a few exceptions, solutions to the \scft{} equations are generally determined using numerical methods. For an ordered phase, numerically solving the \scft{} equations could be conveniently done within one unit cell of the structure. However, the resulting free energy will be dependent on the unit-cell parameters, and therefore, minimization over these parameters must also be performed. We employ Anderson mixing~\cite{thompson2004improved} to iteratively solve the \scft{} equations, and gradient descent to find the optimal lattice parameters. The pseudospectral method~\cite{Tzeremes2002,Rasmussen2002} is employed within each iteration to solve the diffusion equations, Eqs.~\eqref{eq_mod_diff_p1}, and \eqref{eq_mod_diff_p3}. Calculations proceed until the maximum difference in the free energy density between three consecutive iterations is less than $10^{-6}$. The equilibrium phase is then determined by a comparison of the free energies of the candidate structures. In this study, we consider the disordered (\DIS{}), lamellar (\LAM{}), and hexagonally-packed cylindrical (\HEX{})  phases alongside the bicontinuous \dg{} (G), \dd{} (DD), and \pn{} (P) phases as the candidate structures. In the grand canonical ensemble formulation of the theory, the chemical potential  $\mu_2$ is the control parameter. The chemical potentials can be related to the average volume fractions of either polymer types via,
\begin{equation}
\avgpc{2} = \zpot{2}\Nt{2}Q_2 = 1 - \avgpc{1} = 1 - \Nt{1}Q_1,
\end{equation}
where the last two equalities were obtained from the incompressibility condition, $\avgpc{1}+\avgpc{2}=1$, and the condition $\mu_1 = 0$.

\section{Results and Discussions}
\label{sec_results}
In this section, we present our results on the phase behaviour of binary blends composed of gyroid-forming \dbc{}s and \hpl{} copolymers. We use the subscripts $\ig$ and $\ihl$ to denote quantities associated with the gyroid-forming and \hpl{} copolymers, respectively. The relative lengths of the two polymers are characterized by the parameter $\alpha = N_\ihl/N_\ig$. For the \hpl{} copolymer, we set its A-block volume fraction to be $\fhl = N_{\mm{A},\ihl}/N_\ihl = 0.95$. 

\begin{figure}[t]
\includegraphics[width=\imagewidth]{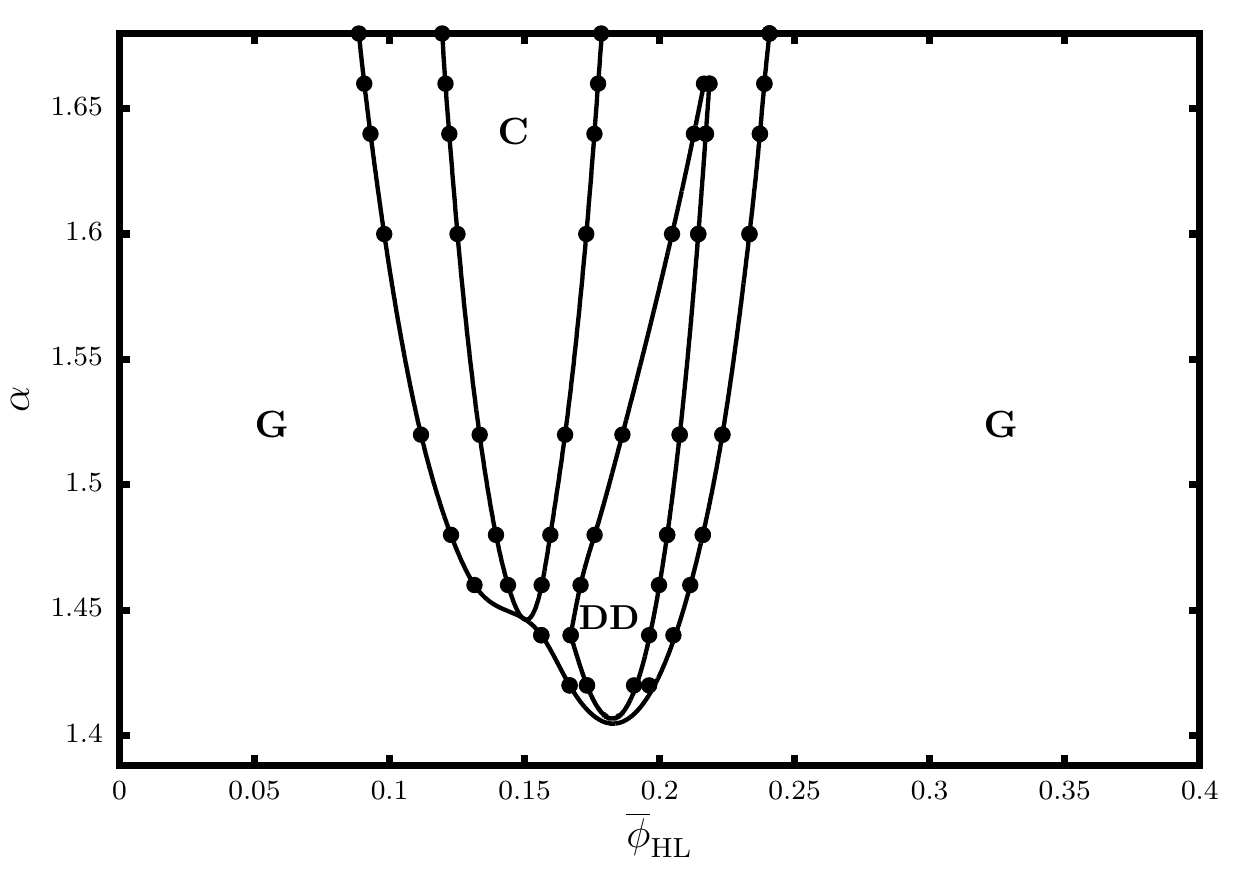}
\caption{Phase diagram of the binary mixture with $\xAB{}\Nt{\mm{\ig}} = 25$, $\fA{\ig} = 0.33$, and $\fA{\ihl} = 0.95$ in the $\avgpc{\ihl}$-$\alpha$ plane. The labels, \HEX, \DG{}, and \DD{}, correspond to the hexagonal, \dg{}, and \dd{} phases, respectively. Unlabeled areas denote regions of phase co-existence between the two bordering phases. The circular points represent the numerically computed data, while the lines serve as guide for the eyes.} 
\label{fig_PD_fa33} 
\end{figure}

We first examine the effect of the blend composition ($\avgpc{\ihl}$) and the ratio of total polymer lengths ($\alpha$) on the phase behaviour of the blends. Figure.~\ref{fig_PD_fa33} shows the phase diagram in the $\vfhl$-$\alpha$ plane for the case of $\fdg = 0.33$ and $\xAB N_\ig = 25$. In all the phase diagrams, the single-phase regions are labeled their phases, whereas the two-phase coexistence regions are left unlabeled.  As shown in Figure~\ref{fig_PD_fa33}, the \DG{} phase is the only equilibrium structure when $\vfhl = 0$ as expected. A new phase emerges once a sufficient amount of \hpl{} copolymers are added to the system. For $1.4 \leq \alpha \leq 1.45$, the \DG{} phase is replaced by the \DD{} phase, whereas for $\alpha \geq 1.45$, hexagonally-packed cylinders are predicted to appear first. The cylindrical phase subsequently transforms to the \DD{} phase upon further increasing $\vfhl$ except near the largest values of $\alpha$ shown. In all cases, the equilibrium morphology reenters to the \dg{} phase once the additive concentration exceeds $\vfhl = 0.25$. 

There are a number of differences in the phase behaviours of binary blends of diblock copolymers mixed with homopolymers (AB/A) or \hpl{} diblock copolymers (AB/AB). First of all, the emergence of the \DD{} phase in \ahp+\adbc{} blends has been predicted to occur for the cases where the \hp{}s are shorter than the \DG-forming \dbc{}s ($\alpha < 1$)~\cite{Martinez-Veracoechea2007,Martinez-Veracoechea2009,Martinez-Veracoechea2009_PLNM, Martinez-Veracoechea2016}. On the other hand, it is clear from the Figure~\ref{fig_PD_fa33} that the \DD{} phase could only be accessed when the homopolymer-like copolymers are longer than the \DG-forming diblock copolymers ($\alpha > 1$). Secondly, the \DD{} or \PN{} phase in the AB/A system is always found at blend compositions where the system is close to becoming saturated with \hp{}s, such that further increasing the \hp{} concentration would cause macrophase separation between the \ahp-rich disordered phase and the ordered \DD{} or \PN{} phase. On the other hand, the AB/AB system could continue to accept additional \hpl{} copolymers after reaching the \DD{} phase, eventually reentering to the \dg{} phase. It is worthwhile to note that the subsequent re-entrance to the \dg{} phase, as well as the phase transition sequence \DG{} $\leftrightarrow$ \DD{} $\leftrightarrow$ \DG{} have been observed experimentally in ternary mixtures of simple SI \dbc{}s~\cite{takagi2019bicontinuous}. 

\begin{figure}[t]
\includegraphics[width=\columnwidth]{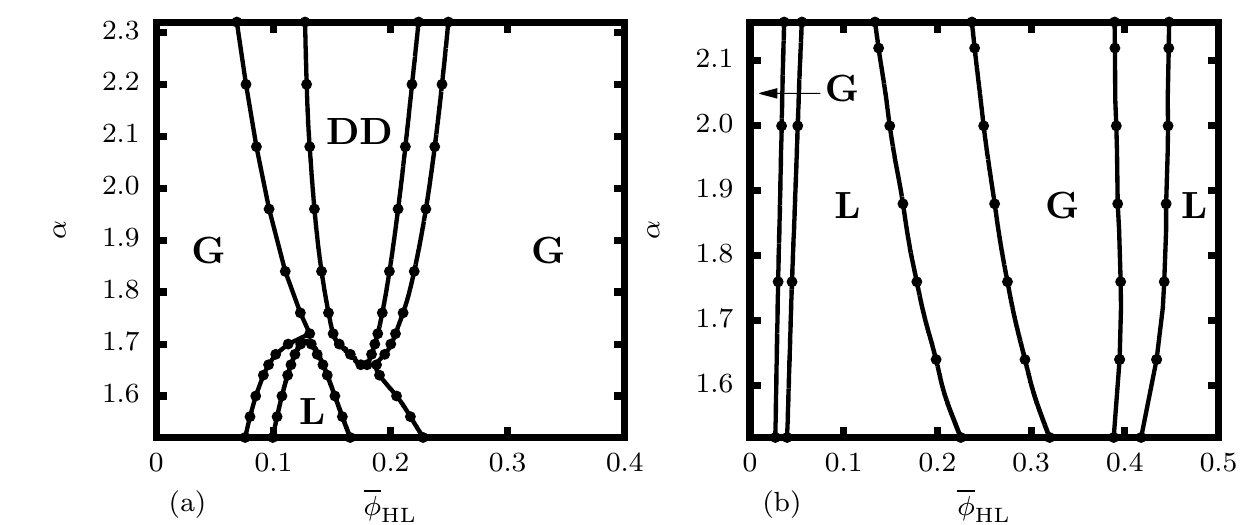}
\caption{Phase diagrams of the binary mixture with $\xAB{}\Nt{\mm{\ig}} = 25$ and $\fA{\ihl} = 0.95$ in the $\avgpc{\ihl}$-$\alpha$ plane. The A-volume fraction of the \DG{}-forming copolymers is set to $\fA{\ig} = 0.34$ in (a) and $\fA{\ig} = 0.35$ in (b). The labels, \LAM, \DG{}, and \DD{}, correspond to the lamellar, \dg{}, and \dd{} phases, respectively.} 
\label{fig_PD_fa34} 
\end{figure}
We next examine the effects of the composition of the gyroid-forming \dbc{}s . Figure~\eqref{fig_PD_fa34} presents phase diagrams in the $\vfhl$-$\alpha$ plane with $\xAB N_\ig = 25$ for two gyroid-forming diblock copolymers with $\fdg = 0.34$ and $\fdg = 0.35$. By design, the equilibrium structure is the \dg{} phase for both diblock copolymers in the absence of the \hpl{} additives ($\vfhl = 0$). It is interesting and surprising that the phase behaviours of these two seemingly similar diblock copolymers deviate significantly from each other when \hpl{} additives are introduced. In the case of $\fdg = 0.34$ (Figure~\ref{fig_PD_fa34} (a)), the lamellar phase emerges for $\alpha \leq 1.65$ and the \DD{} phase appears for $\alpha \geq 1.7$ as $\vfhl$ is increased from 0. At larger $\vfhl$, a re-entrant transition to the gyroid phase occurs, similar to the case of $\fdg=0.33$ shown in Figure~\ref{fig_PD_fa33}. The phase behaviour changes drastically when $\fdg$ is increased slightly to $\fdg = 0.35$. As shown in Figure~\ref{fig_PD_fa34} (b), in this case the intervening \DD{} phase is replaced by the lamellae, resulting in a phase sequence of \DG$\leftrightarrow$\LAM$\leftrightarrow$\DG$\leftrightarrow$\LAM{} when $\vfhl$ is increased. The phase diagrams shown in Figures~\ref{fig_PD_fa33} and \ref{fig_PD_fa34} demonstrate clearly that the composition of the \DG-forming diblock copolymers has a drastic effect on the relative stability of the \bc{} phases, in agreement with the result observed in the \ahp+\adbc{} blends~\cite{Martinez-Veracoechea2007,Martinez-Veracoechea2009,Martinez-Veracoechea2009_PLNM, Martinez-Veracoechea2016}. 

\begin{figure}[t]
\includegraphics[width=\imagewidth]{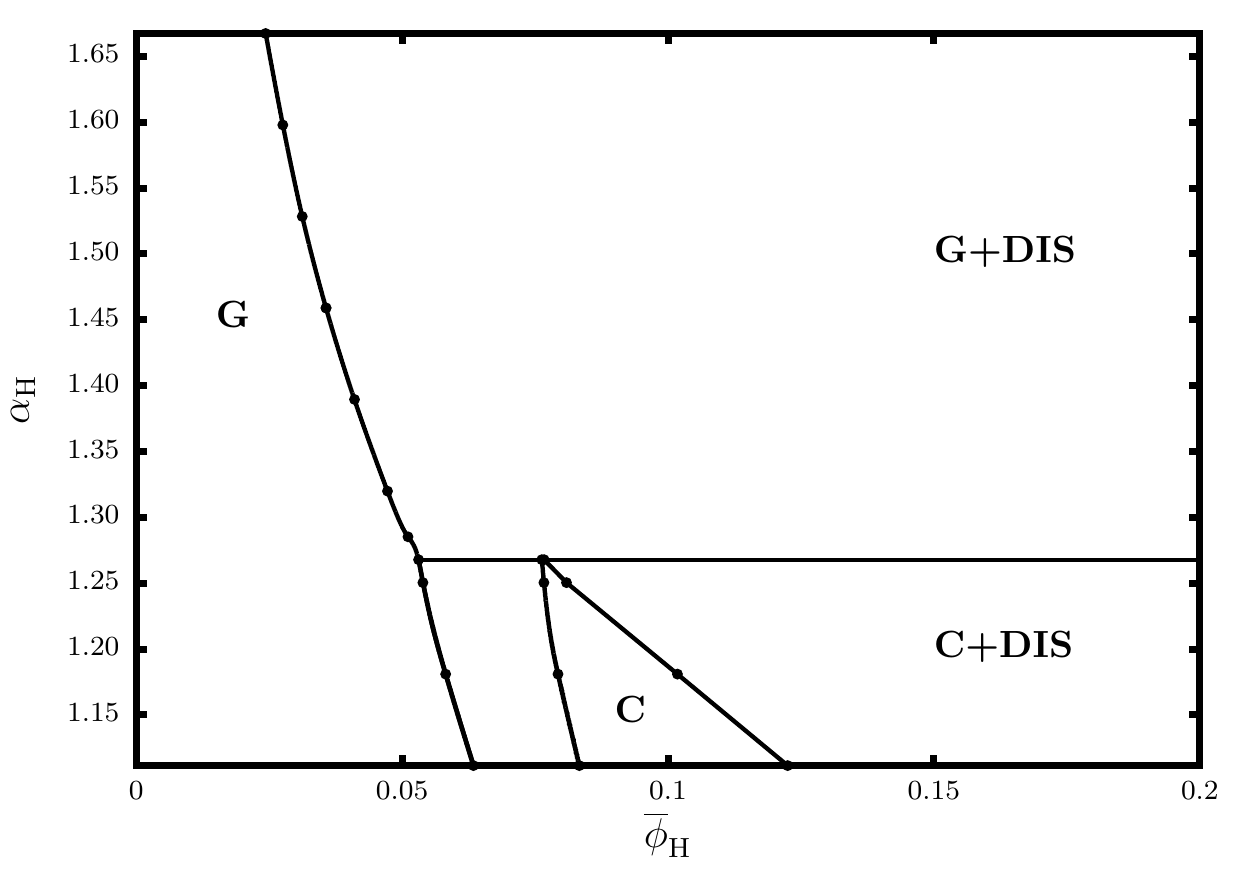}
\caption{Phase diagram of the binary blend consisting of A-\hp{}s and \DG{}-forming \dbc{}s in the $\avgpc{\ihp}$-$\alphahp$ plane, with $\fA{\ig} = 0.33$ and $\xAB{}\Nt{\mm{\ig}} = 25$. The labels, \DIS, \HEX, and \DG{}, correspond to the disordered, hexagonal, and \dg{} phases, respectively.} 
\label{fig_PD_H_DB} 
\end{figure}
We next compare the phase behaviours of binary blends of gyroid-forming diblock copolymers mixed with homopolymers or \hpl{} diblock copolymers. Figure~\ref{fig_PD_H_DB} presents the phase diagram of for the diblock copolymer/homopolymer blends as a function of the homopolymer concentration $\vfhp$ and the homopolymer-\dbc{} length ratio $\alphahp = N_\ihp/N_\ig$, with $\fdg = 0.33$, and $\xAB N_\ig = 25$. In order to compare with the results for the binary AB/AB blends, we focus on the th case of $\alphahp > 1$. It is noted that \DG{}-forming \dbc{}s mixed with shorter \hp{}s have been extensively studied by Matsen~\cite{matsen1995stabilizing,matsen1995phase} and by the Escobedo group~\cite{Martinez-Veracoechea2007,Martinez-Veracoechea2009,Martinez-Veracoechea2009_PLNM, Martinez-Veracoechea2016}. Figure~\ref{fig_PD_H_DB} shows that the phase behaviour of the binary AB/A blend with longer homopolymers is relatively simple. As the homopolymer concentration increases, the parent \dg{} phase transforms to the hexagonally-packed cylinders, which persists until macrophase separation occurs between the cylindrical phase and the \hp{}-rich disordered phase when $\alphahp < 1.25$. Above $\alphahp  = 1.25$, only the \dg{} phase is predicted to occur before the blend becomes saturated with \hp{}s. We find that the value of $\vfhp$ where macrophase separation occurs decreases with increasing $\alphahp$, which is consistent with the fact that the homopolymer solubility decreases when the homopolymers become longer. The absence of the \DD{} phase could be attributed to the low solubility of the longer \hp{}s, {\em i.e.} the blend becomes saturated before reaching the amount of \hp{}s required to sufficiently relieve the packing frustration thus to stabilize the \dd{} phase. It is clear from comparing Figures~\ref{fig_PD_fa33}, and \ref{fig_PD_H_DB} that the $5\%$ replacement of A-monomers with B-monomers in the additives greatly improves its solubility in the blends.

\begin{figure}[t]
\includegraphics[width=\imagewidth]{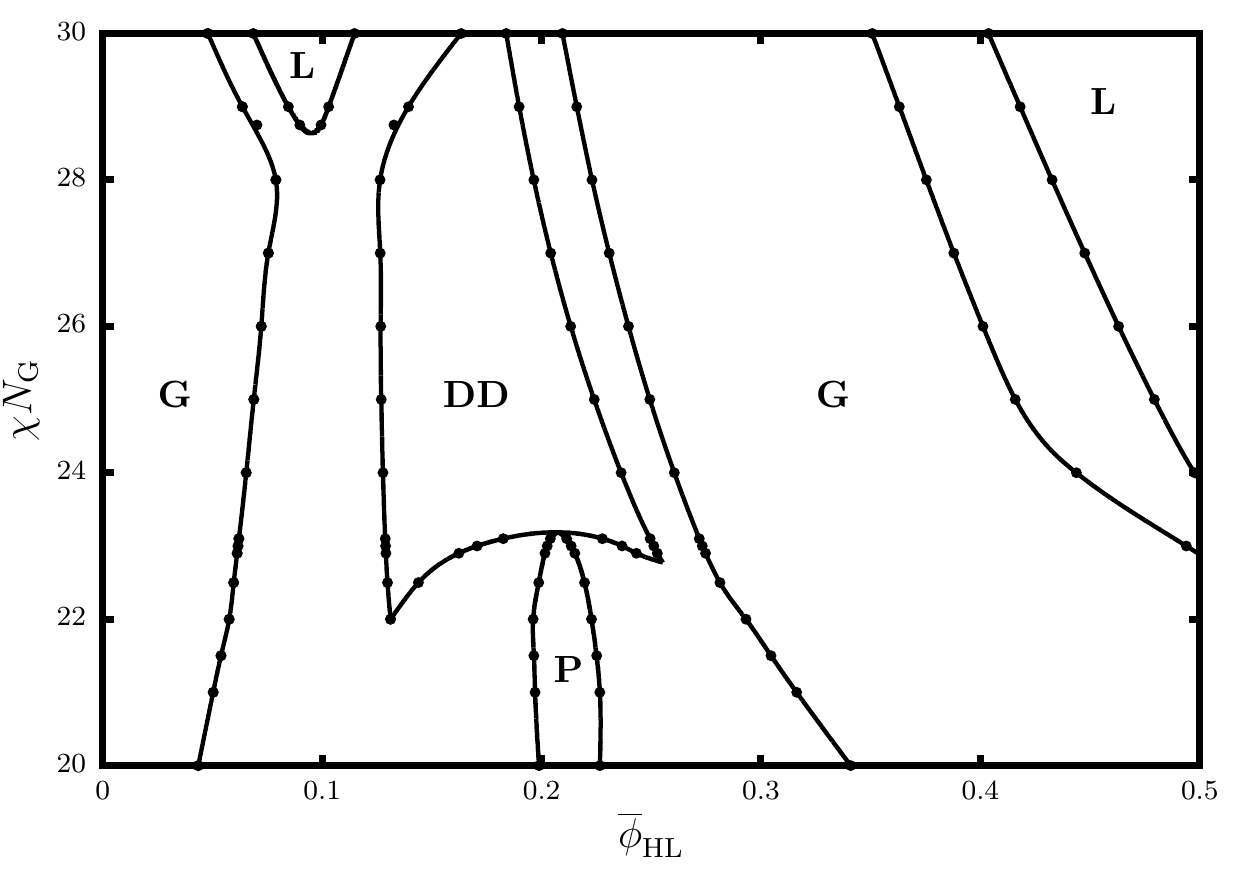}
\caption{Phase diagram of the binary mixture with $\alpha= 2.32$, $\fA{\ig} = 0.34$, and $\fA{\ihl} = 0.95$ in the $\avgpc{\ihl}$-$\xAB{}\Nt{\mm{\ig}}$ plane.  The labels, \LAM, \DG{}, \DD{}, and \PN{} correspond to the lamellar, \dg{}, \dd{}, and \pn{} phases, respectively.} 
\label{fig_PD_chi} 
\end{figure}
We now turn to the effect of the segregation strength on the relative stability of the \bc{} phases. Figure~\ref{fig_PD_chi} presents the phase diagram in the $\vfhl$-$\xAB N_\ig$ plane with $\fdg = 0.34$ and $\alpha = 2.32$. These values of the parameters are chosen to yield the largest range of blend compositions where the \DD{} phase could be accessed. As $\vfhl$ is varied, a phase transition sequence of \DG{}$\rightarrow$\DD$\rightarrow$\DG{} is again predicted. It is interesting to observe that decreasing the segregation strength starting from $\xAB N_\ig = 30$ improves the stability of the \DD{} phase by widening its region of stability. Below $\xAB N_\ig = 23$, a third bicontinuous phase, namely the \PN{} phase, is predicted to appear. This result demonstrates that all three \bc{} structures, {\em i.e.}, the G, DD and P phases, could be accessed by using the \hpl{} diblock copolymers as additives. Furthermore, compared with diblock copolymer/homopolymer blends, the solubility of the \hpl{} copolymers in the blends is greatly enhanced, resulting in a wider range of blend compositions where the emergence of the \DD{} or \PN{} phases is possible. 

To some extent, the improved stability of the \DD{} phase and emergence of the \PN{} phase at lower segregation strengths could have been anticipated. It has been shown that the variations in the mean curvature on the AB interface is greatest for the \PN{} phase, and decreases as one proceeds \PN{}$\rightarrow$\DD$\rightarrow$\DG~\cite{Matsen1994}. Deviations from constant mean curvature leads to excessive interfacial surface area, which is enthalpically penalized. By decreasing the segregation strength, the enthalpic cost for excessive surface area is reduced. In turn, the stability of the \bc{} phases should improve, with improvements being greater for structures having more excessive interfacial area. This makes it more likely for the formation of the \PN{} phase to occur, provided the accompanying entropic gain outweighs the enthalpic penalty. 

Mechanisms responsible for the formation of the \DD{} and \PN{} phases in the current AB/AB blends are related to the roles played by the additives. It is useful to start by reviewing why the the \DD{} and \PN{} phases appear in \ahp+\adbc{} blends. As mentioned earlier, a \bc{} structure with constant mean-curvature will necessarily lead to a difference in thickness or size between the nodes and the struts. This gives rise to packing frustration since polymers have to stretch excessively to occupy the space at the center of the nodes that is entropically unfavourable, or else deform the nodes that is enthalpically unfavourable. Packing frustration could be alleviated through the addition of \hp{}s. The added \hp{}s will aggregate in the nodes, as opposed to the struts, in order to maximize its conformational entropy within the network domain~\cite{Martinez-Veracoechea2016}. In turn, this localization of \hp{}s eliminates the need for \dbc{}s to overly stretch, thus reducing packing frustration. Therefore, as the additive concentration increases, the system proceeds to \bc{} structures with larger nodes, {\em i.e.} \DG$\rightarrow$\DD$\rightarrow$\PN, in order to accommodate more \hp{}s at the nodes.

\begin{figure}[t]
\centering
\includegraphics[width=\imagewidth]{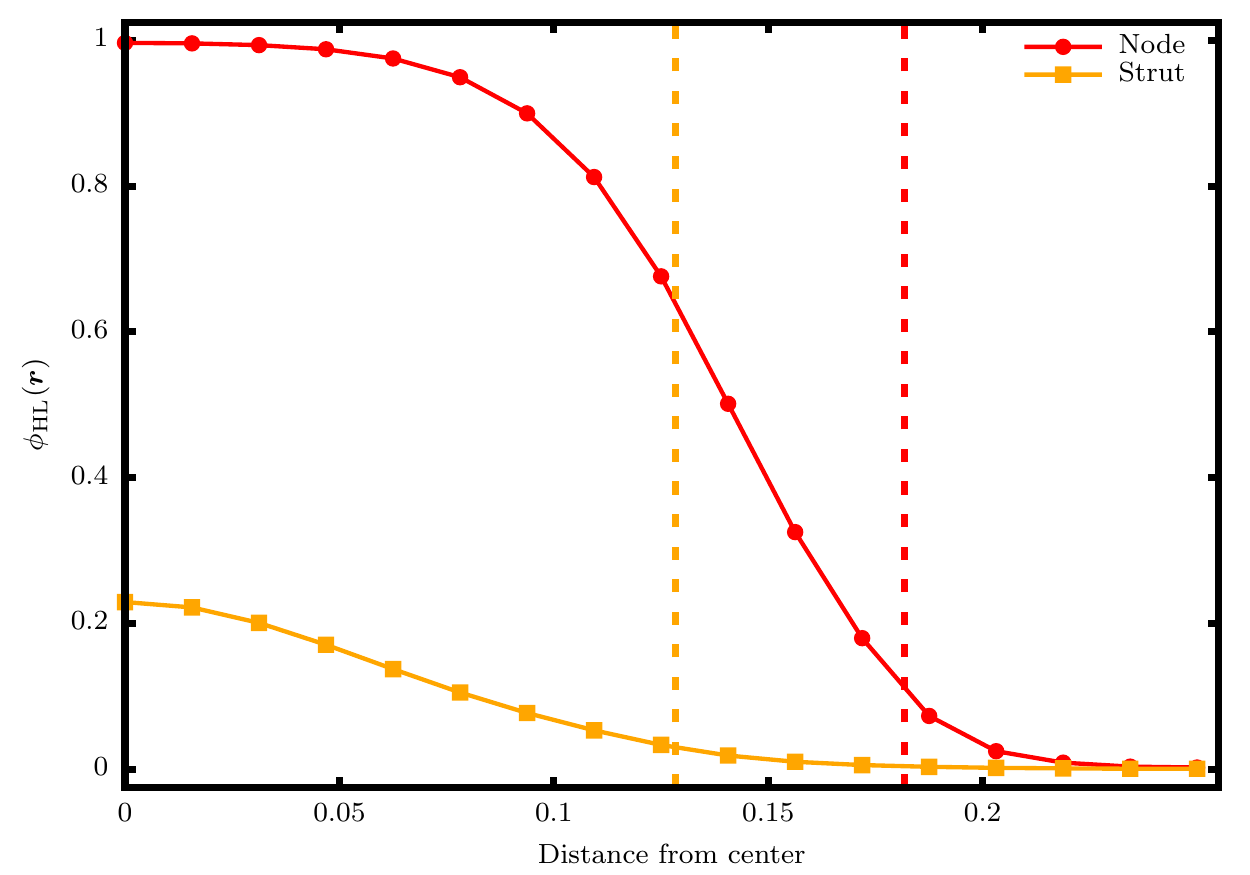}
\caption{Plots of the volume fraction of \hpl{} copolymers $\phi_\mm{\ihl{}}(\vr{})$ as functions of the distance from the center of a node in the [111] direction, and the distance from the center of a strut in the [100] direction for an equilibrium \PN{} phase. The data was taken from an SCFT calculation with $\avgpc{\ihl} = 0.2$, $\fA{\ihl} = 0.95$, $\fA{\ig} = 0.34$, $\xAB N_\ig = 22.5$, and $\alpha = 2.32$. Distances have been rescaled by the length of the unit cell. The dashed lines indicate the positions of the AB interface.} 
\label{fig_nodes_vs_struts_p1} 
\end{figure}
We can further explore these ideas by examining the spatial distribution of the \hpl{} copolymers within the self-assembled structures. In Figure~\ref{fig_nodes_vs_struts_p1}, the additive volume fraction $\phi_\ihl(\vr)$ is plotted as a function of the distance from the center of a node and from the center of a strut halfway between two nodes for a typical equilibrium \PN{} phase. Figure.~\ref{fig_nodes_vs_struts_p1} shows that the concentration of the \hpl{} copolymers is nearly unity near the center of the node. As one moves outwards, and approaches the AB interface, defined customarily as the $\phi_\mm{A}(\vr)=0.5$-isosurface, the \hpl{} copolymer concentration rapidly decreases to zero. The behaviour at the struts is similar in that $\pC{\ihl}$ decays to zero when moving radially outwards from the center. However, the key difference is that the additive concentration near the center of the struts is much lower than that near the center of the nodes. This indicates that the \hpl{} copolymers are primarily confined to the nodes within the A-domain. The results shown in Figure.~\ref{fig_nodes_vs_struts_p1} clearly demonstrate that the \hpl{} diblock copolymers are localized at the center of the nodes, thus acting as space-fillers in the same way as the \hp{}s would behave in \ahp{}+\adbc{} blends, and relieving the packing frustration that hinders the stability of the \DD{} and \PN{} phases. 

\begin{figure}[t]
\centering
\includegraphics[width=\columnwidth]{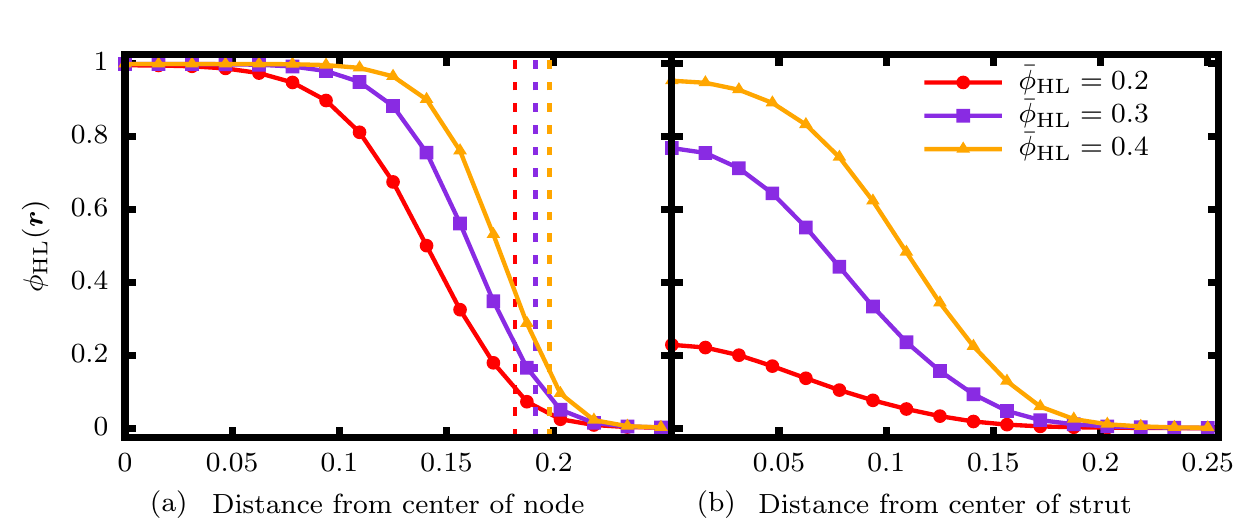}

\caption{Plots of the volume fraction of \hpl{} copolymers $\phi_\mm{\ihl{}}(\vr{})$ as functions of (a) the distance from the center of a node in the [111] direction, and (b) the distance from the center of a strut in the [100] direction for various values of $\avgpc{\ihl}$. The data was taken from SCFT calculations of the \PN{} phase with $\fA{\ihl} = 0.95$, $\fA{\ig} = 0.34$, $\xAB N_\ig = 22.5$, and $\alpha = 2.32$. The dashed lines indicate the positions of the AB interface. Distances have been rescaled by the length of the unit cell. } 
\label{fig_nodes_vs_struts} 
\end{figure}
The spatial distribution of the \hpl{} copolymers depends on their concentration, $\vfhl$, in the system. Figure~\ref{fig_nodes_vs_struts} presents $\phi_\ihl(\vr{})$ as a function of the distance away from the center of a node and from the center of a strut for three values of $\vfhl = 0.2$, $0.3$, and $0.4$. We note that the equilibrium structures at $\vfhl = 0.2$, $0.3$, and $0.4$ are \PN{}, \PN{}+\DG{}, \DG{} phases, respectively. Figure~\ref{fig_nodes_vs_struts} (a) shows that while the AB interface moves slightly further away from the center when $\vfhl$ is increased, the behaviour of $\phi_\ihl(\vr{})$ is qualitatively the same for all three blend compositions. In particular, the central region of the nodes is almost saturated with the \hpl{} copolymers. On the other hand, the distribution of \hpl{} copolymers in the struts behaves quite differently as shown in Figure~\ref{fig_nodes_vs_struts} (b). In particular, the concentration of the \hpl{} copolymers at the center of the struts is quite low at small $\vfhl = 0.2$, and it rapidly increases to a high value when $\vfhl$ is increased. At $\vfhl = 0.4$, the concentration of the \hpl{} copolymers almost reaches 100\% at the center of the struts.  At this point, the additive \hpl{}copolymers are localized at the center of the nodes {\em and} the struts, {\em i.e.}, they are dispersed almost uniformly throughout the network domain. The accumulation of the \hpl{} copolymers in the struts could be the origin of the subsequent transitions from \bc{} structures with more struts per node to ones with less struts per node, {\em i.e.} \PN$\rightarrow$\DD$\rightarrow$\DG. Homopolymer-like copolymers residing in the thinner nodes will experience greater confinement than those located in the nodes, which leads to a loss in conformational entropy. By returning to \bc{} phases with less struts per node, the size of the nodes, and the struts can be made more similar. This should allow \hpl{} copolymers both within the struts, and nodes to better maximize their conformational entropy within the A-domain.

\begin{figure}[t]
\includegraphics[width=\imagewidth]{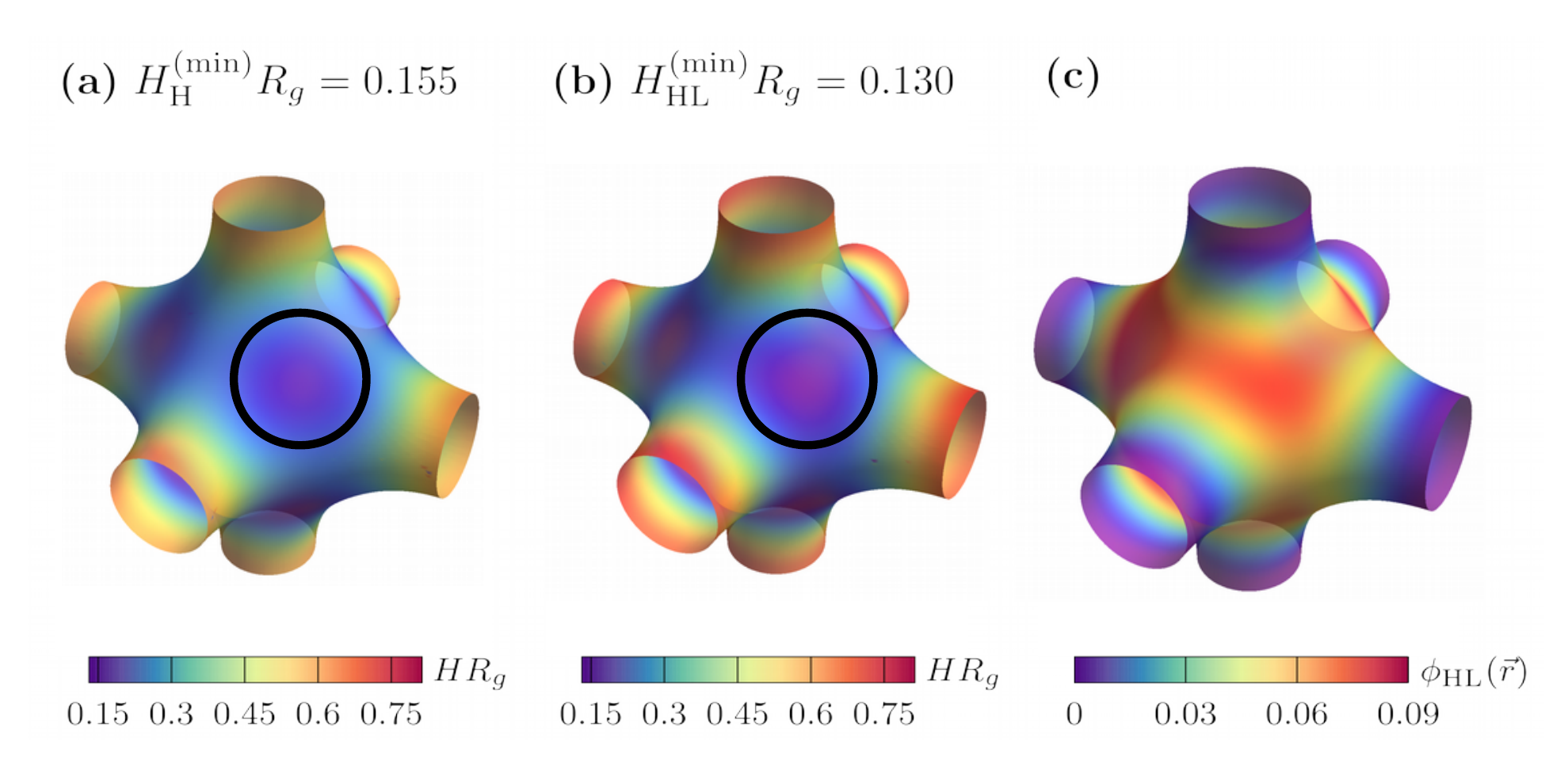}
\centering
\caption{Plots depicting the mean curvature $H$ of the AB-interfacial surface ($\phi_\mm{A}(\vr{})$ = 0.5) for (a) $\avgpc{\ihp} \simeq 0.16$, and (b) $\avgpc{\mm{\ihl}} \simeq 0.16$ with $\fA{\ihl} = 0.95$, $\fA{\ig} = 0.34$, $\xAB N_\ig = 25$, and $\alpha = 2.32$. In (c),  the volume fraction of the \hpl{} copolymers $\phi_{\mm{\ihl}}(\vecr{})$ on the AB interface is shown for the same P phase in (b).} 
\label{fig_homo_DB_curvature_draft} 
\end{figure}
To better understand the enhanced stability of the \bc{} phases in the binary blends, we compare the mean curvature $H$ of the AB interface for the P phase in binary blends of \dbc{}/\hp{} with $\avgpc{\ihp} = 0.16$ (Fig.~\ref{fig_homo_DB_curvature_draft}(a)) and \dbc{}/\hpl{} with $\avgpc{\ihl} = 0.16$ (Fig.~\ref{fig_homo_DB_curvature_draft}(b)). It is known that the packing of the different blocks with different sizes tends to curve the AB interface towards the minority domain in monodispersed \dbc{}s~\cite{Matsen:2002p437,semenov1985contribution}. For the bicontinuous phases, the regions of lowest interface curvature are curved towards the majority B domain, and is therefore entropically unfavourable for the \DG{}-forming DBCPs. The segregation of the \hpl{} DBCPs to the low curvature areas relieves this curvature frustration, resulting in flatter interfaces. This effect is clearly demonstrated by the difference in the minimum value of average curvature $H$ between the cases with \hpl{} and \hp{} additives shown in Figures.~\ref{fig_homo_DB_curvature_draft}(a) and (b). Moreover, the difference in the interfacial curvature between the \hp{}s and \hpl{} copolymers is correlated with the local segregation of the two polymers on the interface as shown in Fig.~\ref{fig_homo_DB_curvature_draft}(c), where the concentration $\phi_{\mm{\ihl}}(\vr{})$ of the \hpl{} DBCPs on the AB interface is plotted. It is clear that the distribution of the additive copolymers is not spatially uniform in that the two types of polymers are locally segregated on the interface. Furthermore, the regions where the additive concentration is the highest coincides with the regions with lowered curvature. The enhanced stability of the various \bc{} phases in the binary blends containing \hpl{} DBCPs could be attributed to the local regulation of the AB interfaces, {\em i.e.} the flattening of the interface regions at the nodes. As a result of these two types of \hpl{} DBCP segregations, {\it viz.} segregation to the center of the nodes and local segregation on the AB interfaces, the \hpl{} DBCP functions as a ``space filler'' to relieve the packing frustration {\em and} as ``co-surfactant" to regulate the interfacial curvature. These two mechanisms act in tandem resulting in greatly enhanced stability of the different \bc{} phases.

\section{Conclusion}
\label{sec_conclusion}
We have examined the formation of various \bc{} phases in binary mixtures of AB/AB \dbc{}s using the self-consistent field theory. We focused on the binary blends composed of gyroid-forming AB \dbc{}s with $\fA{\ig} = 0.33\,\,\mm{to}\,\,0.35$ mixed with a \hpl{} AB diblock copolymer with $\fA{\ihl} = 0.95$. Phase diagrams of the system have been constructed based on SCFT calculations. The results demonstrated that the double-diamond and \pn{} phases, which are metastable in neat diblock copolymer melts, could be stabilized by adding the \hpl{} diblock copolymers. Furthermore, the range of blend compositions over which the \DD{} and \PN{} phases could be accessed is larger than that of the binary \ahp{}+\adbc{} blends.  When the \hpl{} copolymers are replaced by \hp{}s, the resultant binary AB/A blends do not exhibit the \DD{} and  P phases when the homopolymer is longer than the diblock copolymer. This suggests that the $5\%$ replacement of the A monomers with B monomers in the \hp{} plays an essential role in stabilizing the \DD{} and \PN{} phases. By examining the spatial distribution of the different polymeric species within the self-assembled structures, we find that the \hpl{} copolymers are confined primarily to the nodes, which alleviates the packing frustration. We also find that there is a local segregation of the two types of polymers on the AB interface, which results in beneficial changes in the interfacial curvature, further enhancing the stability of the \bc{} phases. In the current study, we have focused on the effects of the polymer-length ratio, the segregation strength, and the composition of the \DG{}-forming diblock copolymers on the formation of novel \bc{} phases. It is expected that tuning the molecular composition of the \hpl{} additive and, more interestingly, using disperse diblock copolymers with designed dispersity distributions, could provide opportunities to further improve the stability of the \DD{}, or \PN{} phases, or perhaps even to stabilize other exotic \bc{} structures such as the Neovious phase~\cite{stroem1992cubic}. 

\section{Acknowledgments}
This research was supported by a Discovery Grant from the Natural Science and Engineering Research Council (NSERC) of Canada and was enabled in part by support provided by the facilities of SHARCNET (https://www.sharcnet.ca) and Compute Canada (http://www.computecanada.ca). 

\bibliography{new_ref}
\end{document}